\documentclass[conference]{IEEEtran}
\usepackage{graphicx,amsmath,array,subfigure,url,booktabs,xcolor,multirow,color,balance,amsfonts,algorithmic,algorithm,textcomp,stfloats,verbatim,newclude,accents,amssymb,amsthm,cite,placeins,bm}

\newtheorem{theorem}{Theorem}

\begin{document}

\title{Polarforming Design with Phase Shifter Based Polarization Reconfigurable Antennas\\
%\thanks{\textit{(Corresponding authors: Jingze Ding and Rui Zhang.)}}
}

\author{\IEEEauthorblockN{
	Zijian Zhou\IEEEauthorrefmark{2}, 
	Jingze Ding\IEEEauthorrefmark{3}, 
	and Rui Zhang\IEEEauthorrefmark{2}\IEEEauthorrefmark{4}}
\IEEEauthorblockA{\IEEEauthorrefmark{2}School of Science and Engineering, The Chinese University of Hong Kong, 		Shenzhen 518172, China}
\IEEEauthorblockA{\IEEEauthorrefmark{3}School of Electronics, Peking University, Beijing 100871, China}
\IEEEauthorblockA{\IEEEauthorrefmark{4}Department of Electrical and Computer Engineering, National University of Singapore, Singapore 117583}
\IEEEauthorblockA{Email: zijianzhou@link.cuhk.edu.cn, djz@stu.pku.edu.cn, rzhang@cuhk.edu.cn}}

\maketitle

\begin{abstract}
In this paper, we propose a new form of polarization reconfigurable antennas (PRAs) that can form linear, circular, and general elliptical polarizations assisted by phase shifters (PSs).  With PRAs, polarforming is achieved, which enables the antenna to shape its polarization into a desired state for aligning with that of the received electromagnetic (EM) wave or reconfiguring that of the transmit EM wave.  To demonstrate the benefits of polarforming, we investigate a PRA-aided single-input single-output (SISO) communication system equipped with tunable PSs for polarization adaptation.  We characterize the achievable signal-to-noise ratio (SNR) at the receiver as a function of the phase shifts of PS-based PRAs.  Moreover, we develop an alternating optimization approach to maximize the SNR by optimizing the phase shifts at both the transmitter and receiver.  Finally, comprehensive simulation results are presented, which not only validate the effectiveness of polarforming in mitigating the channel depolarization effects, but also demonstrate its substantial performance improvement over conventional systems.
\end{abstract}


\section{Introduction} \label{sec1}
\IEEEPARstart{P}{olarization} characterizes the basic property of an electromagnetic (EM) wave and could potentially triple wireless channel capacity by utilizing six distinct electric and magnetic polarization states existing at any point in space \cite{ref_DZJZ25}.  Nonetheless, classical wireless systems place an overfull focus on optimizing time, frequency, and spatial resources, often neglecting polarization as another dimension, especially in recent promising technologies like movable/fluid antennas \cite{ref_ZMZ24, ref_DZL24, ref_LDZ25, ref_DZJ25, ref_WST25, ref_DZZ25, ref_MWN24, ref_DZS24} and pinching antennas \cite{ref_DSP25}.  Although these systems have been successfully implemented or investigated, they are now reaching their theoretical and practical performance boundaries.  To meet the unprecedented demands for higher data rates and more reliable connections, leveraging polarization diversity in wireless systems is becoming a promising and necessary evolution.

Since the events that cause channel depolarization are random and highly dependent on the surrounding environment, conventional systems with fixed-polarization antennas (FPAs) cannot effectively adapt to channel variations and mitigate channel depolarization.  The dual-polarized antenna (DPA) is the most well-explored technology for leveraging polarization diversity and overcoming the inherent limitations of FPAs \cite{ref_OB23}.  In DPA systems, each antenna has two ports with orthogonal polarization orientations and requires two dedicated radio frequency (RF) chains.  However, the use of double RF chains in DPA systems considerably raises its cost, making it impractical for lightweight, low-complexity wireless devices in many applications such as the Internet of Things \cite{ref_XHL14}.  To reduce the cost of DPAs, polarization reconfigurable antennas (PRAs) have been developed, commonly known as switchable PRAs (SPRAs) due to their capability to switch between predefined polarization states \cite{ref_KRR15}.  The implementation of SPRAs is well-established, relatively low-cost, and benefits from advancements in antenna architecture.  Nevertheless, SPRAs usually support a limited number of predefined polarization states, e.g., two and four.  This limitation constrains the ability of the SPRA system to fully leverage polarization diversity and combat channel depolarization.  An alternative approach is employing polarization agile antennas (PAAs) to overcome the discrete adjustment limitations of SPRAs while maintaining a single RF chain per antenna \cite{ref_KM15, ref_CH24}.  The PAA system enables continuous adjustment of polarization orientation through linearly polarized antennas (LPAs) within a given antenna panel.  However, PAAs can mitigate channel depolarization to some extent; however, the performance is limited by the use of LPAs, instead of circularly polarized antennas (CPAs) \cite{ref_DGM17}.

Motivated by the above, this paper proposes a new type of PRA based on phase shifters (PSs) for enhancing wireless communication performance.  Different from the existing PRAs mentioned above, the PS-based PRA enables adaptive polarization using low-cost PSs integrated into the antenna front-end.  Furthermore, it is shown by simulation that the PS-based PRA outperforms the conventional FPAs and PRAs in adapting to channel fading and combating channel depolarization under typical polarized channel conditions.  More specifically, the main contributions of this paper are summarized as follows.  First, we propose a new paradigm of PS-based PRAs, where each PRA can independently control the phase difference of vertical and horizontal polarizations to form linear, circular, and general elliptical polarizations.  Second, we develop an alternating optimization approach to maximize the receive signal-to-noise ratio (SNR) of a single-input single-output (SISO) system by iteratively optimizing the phase shifts at the transmitter and receiver both equipped with PS-based PRAs.  Finally, we carry out comprehensive simulations to evaluate the performance gains achieved by polarforming over conventional systems under different channel conditions. 

\section{System Model and Problem Formulation} \label{sec2}
\begin{figure*}[!t]
	\centering
	\includegraphics[width=0.8\linewidth]{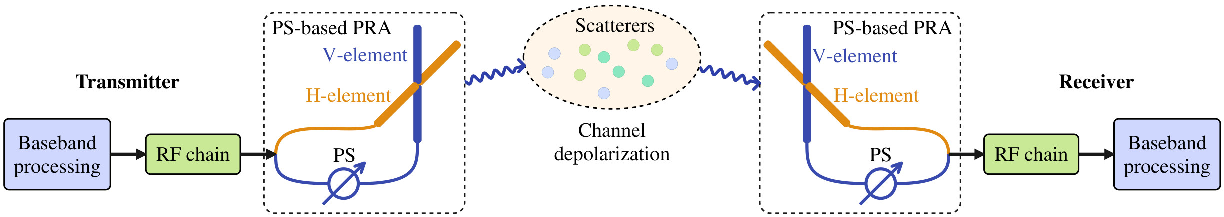}
	\caption{Illustration of the PS-based PRA-enabled wireless communication system.}
	\label{fig_system_model}
\end{figure*}

As shown in Fig.\;\ref{fig_system_model}, we consider a SISO system assisted by PS-based PRAs at the transmitter and receivers.  Each PRA is equipped with only a single RF chain and consists of two orthogonal antenna elements, i.e., V-element for vertical polarization and H-element for horizontal polarization.  While the PS-based PRA in this paper employs two antenna elements, it is worth noting that our model can be extended to accommodate three orthogonal antenna elements by incorporating a PS with the additional antenna element.  A PS is employed to flexibly adjust the phase difference between the two antenna elements, with its tunable range assumed to span from $0$ to $2\pi$.  We denote $\theta$ as the phase shift of the transmit PRA and $\phi$ as the phase shift of the receive PRA.

For the proposed PRA-enabled system, we consider narrow-band quasi-static channels, and the channel response can be expressed as a function of the phase shifts, i.e., $h(\theta, \varphi) \in \mathbb{C}$.  Let $s \in \mathbb{C}$ denote the transmit signal with zero mean and normalized power of one.  Therefore, the received signal of the PRA system can be written as
\begin{equation}
	y(\theta,\phi) = h(\theta,\phi) P_{\rm t} s + z,
\end{equation}
where $P_{\rm t}$ is the transmit power, and  $z$ is the additive white Gaussian noise (AWGN) at the receiver, with an average noise power of $\sigma^2$.  We assume that the noise from antenna elements and PSs is negligible compared to that from the RF chains, as the former is typically much lower than the latter introduced by active RF chain components.

Next, we define transmit and receive polarforming vectors (PFVs) in terms of phase shifts to characterize the polarization of PS-based PRAs as
\begin{equation} \label{PFV}
	{\bf f}(\theta) \triangleq \frac{1}{\sqrt{2}} \begin{bmatrix} 1\\e^{j\theta} \end{bmatrix} \;\text{and}~{\bf g}(\phi) \triangleq \begin{bmatrix} 1\\e^{j\phi} \end{bmatrix},
\end{equation}
respectively.  The factor $\frac{1}{\sqrt{2}}$ ensures normalization to comply with the transmit power constraint.  The transmit and receive PFVs in \eqref{PFV} capture the relationship between phase shifts and polarization configurations of the transmit and receive PRAs.

Using these definitions, the overall channel response can be expressed as $h(\theta, \phi) \triangleq {\bf g}(\phi)^H {\bf P} {\bf f}(\theta)$, where ${\bf P}$ is a polarized channel matrix of two-by-two dimensions, given by ${\bf P} = {\bf \Psi} \odot {\bf H}_{\rm i.i.d.}$ \cite{ref_HCS16}, with
\begin{equation}
	{\bf \Psi} = \frac{1}{\sqrt{\chi + 1}} \begin{bmatrix} 1 & \sqrt{\chi} \\ \sqrt{\chi} & 1 \end{bmatrix}, 
\end{equation}
and $\chi$ represents the inverse cross-polarization discrimination (XPD) indicating the degree of channel depolarization.  The elements of the matrix ${\bf H}_{\rm i.i.d.} \in \mathbb{C}^{2\times 2}$ are i.i.d. and circularly  distributed random variables with equal covariance of $\frac{1}{\sqrt{2}}$ after normalization.

To evaluate the performance limit of the proposed PRA system, we assume that perfect channel state information (CSI) of the polarized channel matrix ${\bf P}$ is available at both the transmitter and receiver\footnote{Acquiring the CSI of the polarized channel matrix is challenging; however, we make this assumption in order to investigate the theoretical performance limit of the system.}.  Under this assumption, both transmit and receive PRAs can adjust antenna polarization to optimize the system performance.  Thus, the SNR of the PRA system at the receiver is given by
\begin{equation} \label{snr}
	\gamma(\theta, \phi) = \frac{|h(\theta,\phi)|^2 P_{\rm t}}{\sigma^2},~\theta,\phi\in[0,2\pi].
\end{equation}

In this paper, we aim to maximize the rate of the PRA system by jointly optimizing the transmit and receive phase shifts, $\theta$ and $\phi$, subject to the phase shift constraint.  Since the rate is given by $R(\theta,\phi) = \log_2(1 + \gamma(\theta,\phi))$, it is equivalent to maximizing the SNR of the considered system.  Accordingly, the optimization problem can be formulated as
\begin{subequations}
\label{opt_problem}
\begin{align}
		\max_{\theta, \phi} ~ & \gamma(\theta,\phi)  \\
		{\rm s.t.} ~ & ~ \theta \in [0,2\pi], \label{cons_theta} \\
		& ~ \phi \in [0,2\pi]. \label{cons_phi}
\end{align}
\end{subequations}
Problem \eqref{opt_problem} is difficult to solve because the objective function is non-concave with respect to $\theta$ and $\phi$ due to the exponential expression in \eqref{PFV}.  Besides, the coupling between the optimization variables $\theta$ and $\phi$ adds to the complexity of the problem.  To address these challenges, we will develop an alternating optimization approach, which iteratively optimizes one of the variables while keeping the other fixed, thereby simplifying the problem into more tractable subproblems.

\section{Proposed Solution} \label{sec3}
In this section, we propose an efficient algorithm to maximize the SNR in \eqref{snr} based on the alternating optimization approach.  

First, we consider that the receive PRA is flexible and the transmit PRA is fixed at a given polarization configuration.  In the subproblem, the polarized channel matrix reduces to an effective polarized channel vector, expressed as ${\bf b}\triangleq {\bf P} {\bf f}(\theta)$.  By denoting ${\bf B} = {\bf b}{\bf b}^T \in \mathbb{C}^{2\times 2}$, problem \eqref{opt_problem} is rewritten as
\begin{subequations}
\label{opt_problem1}
\begin{align}
		\max_{\phi} ~ & {\bf g}(\phi)^H {\bf B} {\bf g}(\phi)  \\
		{\rm s.t.} ~ & ~ \eqref{cons_phi}.
\end{align}
\end{subequations}
In this subproblem, the objective function is with respect to the phase shift $\phi$ of the receive PRA.  By observing that the matrix ${\bf B}$ in \eqref{opt_problem1} is Hermitian, we introduce the following theorem to address problem \eqref{opt_problem1}.

\begin{theorem} \label{theorem1}
	Let ${\bf W}$ be a $2\times 2$ Hermitian matrix, and define ${\bf p}(\psi) = [1, e^{j\psi}]^T$.  The optimal angle $\psi$ that maximizes the quadratic function ${\bf p}(\psi)^H {\bf W} {\bf p}(\psi)$ subject to $\psi \in [0,2\pi]$ is given by
	\begin{equation} \label{opt_psi}
		\psi^\star = \angle{[{\bf W}]_{21}}.
	\end{equation}
\end{theorem}

\begin{IEEEproof}
	Consider a Hermitian matrix ${\bf W}$ expressed as
	\begin{equation}
		{\bf W} = \begin{bmatrix}
			a & c^* \\
			c & d
		\end{bmatrix},
	\end{equation}
	where $a,d\in\mathbb{R}$ and $c\in\mathbb{C}$. The objective function can be rewritten as
	\begin{align} \label{psi_W_psi}
		{\bf p}(\psi)^H {\bf W} {\bf p}(\psi) &= ce^{-j\psi} + c^* e^{j\psi} + a + d \nonumber\\
		&= 2 |c| \cos(\psi-\angle{c}) + a + d.
	\end{align}

	To maximize the objective function ${\bf p}(\psi)^H {\bf W} {\bf p}(\psi)$, the cosine term in \eqref{psi_W_psi} must attain its maximum value of one for $\psi \in [0,2\pi]$.  This occurs when $\psi - \angle{c} = 0$, which implies $\psi^\star = \angle{[{\bf W}]_{21}}$.  Thus, the optimal angle in \eqref{opt_psi} is achieved, and the proof of the theorem is complete.
\end{IEEEproof}

According to Theorem \ref{theorem1}, the optimal phase shift $\phi$ to maximize the objective function ${\bf g}(\phi) {\bf B} {\bf g}(\phi)$ in \eqref{opt_problem1} is given by
\begin{equation} \label{opt_phi}
\phi^\star = \angle{\left[ {\bf P} {\bf f}(\theta) {\bf f}(\theta)^H {\bf P}^H \right]_{21}}.
\end{equation}
The phase shift given in \eqref{opt_phi} represents the optimal solution for polarforming applied at the receiver with an FPA at the transmitter, referred to as receive polarforming, which will be further analyzed in Section \ref{sec4}.

For given $\phi$, we denote ${\bf D} = {\bf P}^H{\bf g}(\phi) {\bf g}(\phi)^H{\bf P}\in \mathbb{C}^{2\times 2}$.  In this subproblem, problem \eqref{opt_problem} can be rewritten as
\begin{subequations}
\label{opt_problem2}
\begin{align}
		\max_{\theta} ~ & {\bf f}(\theta)^H {\bf D} {\bf f}(\theta)  \\
		{\rm s.t.} ~ & ~ \eqref{cons_theta}.
\end{align}
\end{subequations}
To maximize the objective function in problem \eqref{opt_problem2}, it is equivalent to optimizing the phase shift of the transmit PRA.  Thus, the optimal phase shift $\theta$ to maximize the objective function ${\bf f}(\theta)^H {\bf D} {\bf f}(\theta)$ is given by
\begin{equation} \label{opt_theta}
\theta^\star = \angle{\left[ {\bf P}^H {\bf g}(\phi) {\bf g}(\phi)^H {\bf P} \right]_{21}}.
\end{equation}
It can be inferred that for a fixed receive PRA, the optimal solution for the transmit PRA is achieved by aligning the phase shift through phase precoding when full CSI is available at the transmitter.  Actually, the phase shift given in \eqref{opt_phi} provides the optimal solution for adjusting the transmit PRAs with an FPA at the receiver, called transmit polarforming, which will be evaluated in Section \ref{sec4}.

\begin{algorithm}[!t]
	\caption{Proposed Solution for Solving Problem \eqref{opt_problem}}
	\label{alg1}
	\footnotesize
	\renewcommand{\algorithmicrequire}{\textbf{Input:}}
	\renewcommand{\algorithmicensure}{\textbf{Output:}}
	\begin{algorithmic}[1]
		\REQUIRE $P_{\rm t}$, $\sigma$, ${\bf P}$, $\epsilon$.
		\ENSURE $\theta$, $\phi$.
		\STATE Initialize $\theta^{(0)} = 0$ and $\phi^{(0)} = 0$.
		\FOR{$i=1\rightarrow I_{\max}$}
		\STATE Update $\theta^{(i)} \leftarrow \theta^{(i-1)}$ and $\phi^{(i)} \leftarrow \phi^{(i-1)}$.
		\STATE Calculate ${\bf B} = {\bf P}{\bf f}(\theta) {\bf f}(\theta)^H{\bf P}^H$.
		\STATE Update $\phi^{(i)}$ according to \eqref{opt_phi}.
		\STATE Calculate ${\bf D} = {\bf P}^H{\bf g}(\phi) {\bf g}(\phi)^H{\bf P}$.
		\STATE Update $\theta^{(i)}$ according to \eqref{opt_theta}.
		\IF{Increase of the SNR in \eqref{snr} is below $\epsilon$}
		\STATE Break.
		\ENDIF
		\ENDFOR
		\STATE Set $\theta = \theta^{(i)}$.
		\STATE Set $\phi = \phi^{(i)}$.
		\RETURN $\theta$, $\phi$.
	\end{algorithmic}
\end{algorithm}

Based on the above, Algorithm \ref{alg1} provides an iterative method to solve problem \eqref{opt_problem}.  It initializes the variables $\theta^{(0)} = 0$ and $\phi^{(0)} = 0$, and iterates until a maximum number of iterations $I_{\max}$ is reached.  In each iteration, the algorithm computes the matrix products in \eqref{opt_phi} and \eqref{opt_theta}, and subsequently updates the phase shifts $\theta^{(i)}$ and $\phi^{(i)}$.  The process continues until the SNR increase falls below $\epsilon$ or the maximum number of iterations is reached.  The final values of $\theta^{(i)}$ and $\phi^{(i)}$ are returned as the solution.  The convergence of Algorithm \ref{alg1} is guaranteed since the alternating optimization of variables $\theta$ and $\phi$ results in a non-decreasing SNR in each iteration, which is bounded below a finite value.  Moreover, the convergence behavior will be validated through simulations in Section \ref{sec4}.

\section{Simulation Results} \label{sec4}
We carry out extensive simulations to evaluate the performance gains of the proposed system over conventional/benchmark systems and validate the effectiveness of the proposed algorithm.

In the simulations, we assume that the PSs for polarforming can be flexibly adjusted within the range of $0$ to $2\pi$.  We focus on the Rayleigh fading channel, which is also considered in \cite{ref_KM15}.  Unless otherwise stated, the inverse XPD $\chi$ is set to $0.2$ as in \cite{ref_CH24}.  Due to the normalization of the channel matrix, the average SNR of the proposed system is only determined by the transmit power $P_{\rm t}$ and the noise power $\sigma^2$ at the receiver, i.e., $\frac{P_{\rm t}}{\sigma^2}$.  Furthermore, the convergence thresholds for the relative increment of the objective function are set to $\epsilon = 10^{-3}$ for Algorithm \ref{alg1}.  The maximum number of iterations is limited to $I_{\max} = 20$ to ensure efficient convergence within a reasonable computational cost.  To ensure the robustness of the simulations, the results are obtained by averaging over $10^4$ independent Monte Carlo channel realizations.

\begin{figure}[!t]
	\centering
	\includegraphics[width=0.7\linewidth]{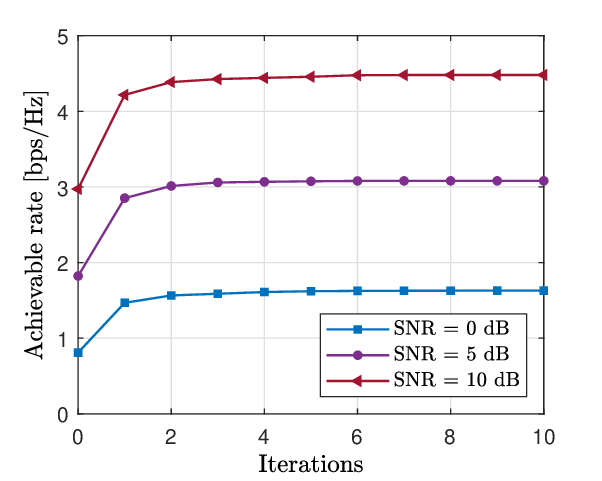}
	\caption{Convergence behavior of Algorithm \ref{alg1}.}
	\label{fig_conv}
\end{figure}

First of all, the convergence behavior of the proposed algorithm for different values of SNR is illustrated in Fig.\;\ref{fig_conv}.  The results demonstrate that the achievable rate consistently increases and reaches a maximum value after six iterations for all values of SNR, which validates the convergence analysis in Sections \ref{sec3}.

In comparison to the proposed scheme marked by ``Polarforming'' with PS-based PRAs, four benchmark schemes are considered, defined below.
\begin{itemize}
	\item \textbf{SPRA}: This system is assisted with SPRAs at the transmitter and/or receiver, where each SPRA is connected to a single RF chain and can switch among two polarization states, i.e., left-handed circular polarization and right-handed circular polarization \cite{ref_KRR15}.  In this scheme, the transmit and receive polarization vectors are given by ${\bf p}_{\rm t} \in \left\{ \frac{1}{\sqrt{2}}[1,j]^T, \frac{1}{\sqrt{2}}[1,-j]^T \right\}$ and ${\bf p}_{\rm r} \in \left\{ [1,j]^T, [1,-j]^T \right\}$, respectively.

	\item \textbf{PAA}: This system is equipped with PAAs at the transmitter and/or receiver, where each PAA can dynamically adjust its polarization angle within the range of $0$ to $2\pi$.  The polarization vectors are defined as ${\bf p}_{\rm t} = [\cos\alpha, \sin\alpha]^T$ and ${\bf p}_{\rm r} = [\cos\beta, \sin\beta]^T$ \cite{ref_KM15, ref_CH24}, with $\alpha$ and $\beta$ being polarization angles at the transmitter and receiver, respectively.
	
	\item \textbf{CPA}: The polarization of the transmit and receive antennas is fixed to left-handed circular polarization, with the polarization vectors given by ${\bf p}_{\rm t} = \frac{1}{\sqrt{2}}[1,j]^T$ for the transmitter and ${\bf p}_{\rm r} = [1,j]^T$ for the receiver.
	
	\item \textbf{LPA}: The polarization of the transmit and receive antennas is fixed to vertical polarization, with the polarization vectors given by ${\bf p}_{\rm t} = [1,0]^T$ for the transmitter and ${\bf p}_{\rm r} = [1,0]^T$ for the receiver.
\end{itemize}
For the above benchmark schemes, the polarization vector at the transmitter is normalized to satisfy the transmit power constraint, but normalization at the receiver is not necessary.

\begin{figure}[!t]
	\centering
	\subfigure[With a single-LPA at the receiver]{
		\centering\label{fig_txp_LPA}
		\includegraphics[width=0.7\columnwidth]{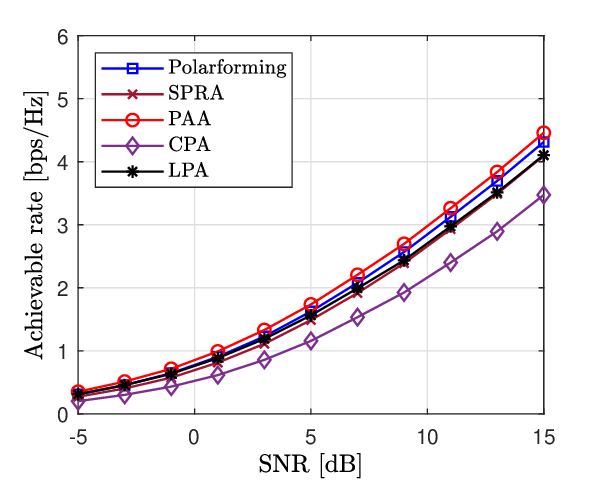}}
	\subfigure[With a single-CPA at the receiver]{
		\centering\label{fig_txp_CPA}
		\includegraphics[width=0.7\columnwidth]{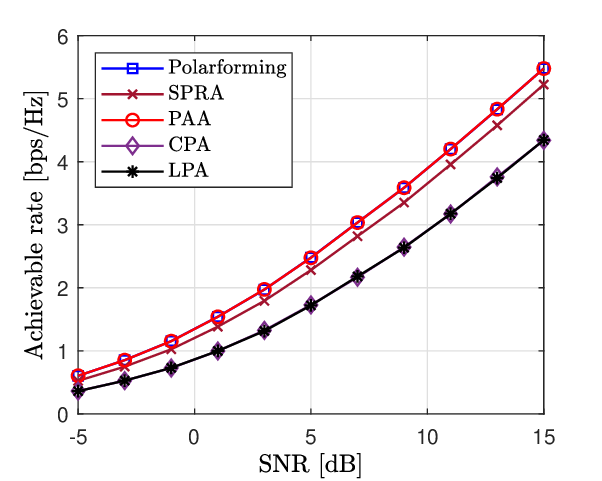}}
	\caption{Achievable rate of transmit polarforming versus SNR when it is applied at the transmitter only.}
	\label{fig_txp}
\end{figure}

Fig.\;\ref{fig_txp} shows the performance of transmit polarforming in the SISO system when it is applied at the transmitter only.  In transmit polarforming, an LPA or CPA is deployed at the receiver, while a PS-based PRA is deployed at the transmitter.  The optimal solution for transmit polarforming is based on \eqref{opt_theta} as discussed in Section \ref{sec3}.  From Fig.\;\ref{fig_txp_LPA}, the schemes using antennas with a single element, such as the PAA and LPA schemes, perform better than the proposed and CPA schemes, respectively.  However, this advantage diminishes when a CPA is used at the receiver, as shown in Fig.\;\ref{fig_txp_CPA}.  This occurs because the antennas with dual elements lose their advantage under the transmit power constraint. Additionally, the schemes using a single element achieve better performance when the inverse XPD is set to $\chi = 0.2$, which corresponds to weak channel depolarization.

\begin{figure}[!t]
	\centering
	\subfigure[With a single-LPA at the transmitter]{
		\centering\label{fig_rxp_LPA}
		\includegraphics[width=0.7\columnwidth]{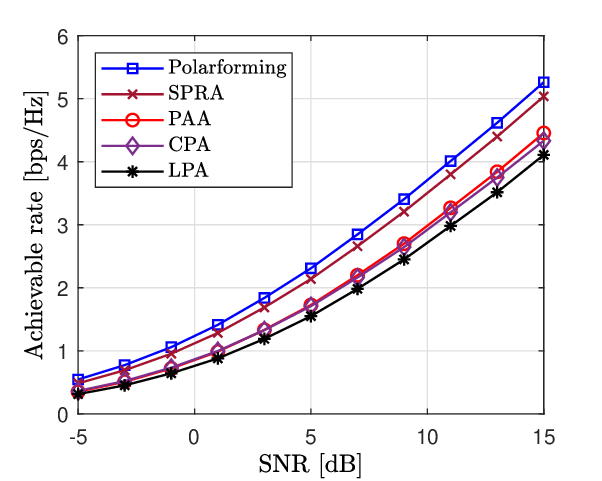}}
	\subfigure[With a single-CPA at the transmitter]{
		\centering\label{fig_rxp_CPA}
		\includegraphics[width=0.7\columnwidth]{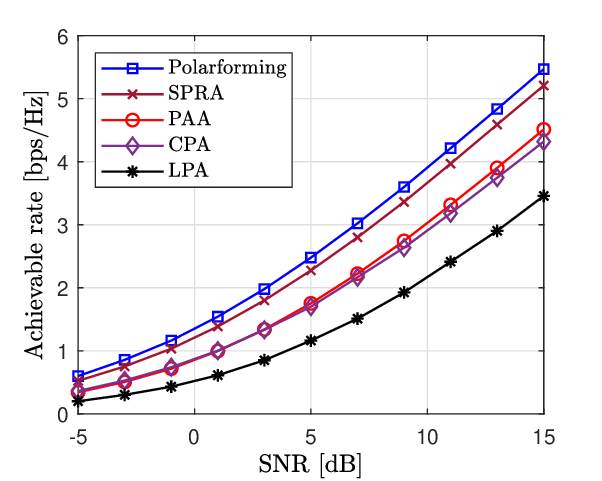}}
	\caption{Achievable rate of receive polarforming versus SNR when it is applied at the receiver only.}
	\label{fig_rxp}
\end{figure}

Fig.\;\ref{fig_rxp} depicts the performance of receive polarforming in the SISO system when it is applied at the receiver only.  In receive polarforming, an LPA or CPA is deployed at the transmitter, and a PS-based PRA is deployed at the receiver.  The optimal solution for receive polarforming is based on \eqref{opt_phi} as discussed in Section \ref{sec3}.  Receive polarforming generally achieves higher rates than transmit polarforming, as shown in Figs. \ref{fig_rxp_LPA} and \ref{fig_rxp_CPA}.  Moreover, the antennas with dual elements outperform those with a single element, such that the proposed scheme surpasses the PAA scheme, and the CPA scheme outperforms the LPA scheme. Notably, flexible-polarization antennas consistently perform better than FPAs, i.e., the proposed and SPRA schemes outperform the CPA scheme, and the PAA scheme outperforms the LPA scheme.  The results in these figures demonstrate that the proposed polarforming scheme generally achieves better performance as compared to benchmark schemes, particularly at the receiver, where it outperforms the PAA scheme.

\begin{figure}[!t]
	\centering
	\includegraphics[width=0.7\linewidth]{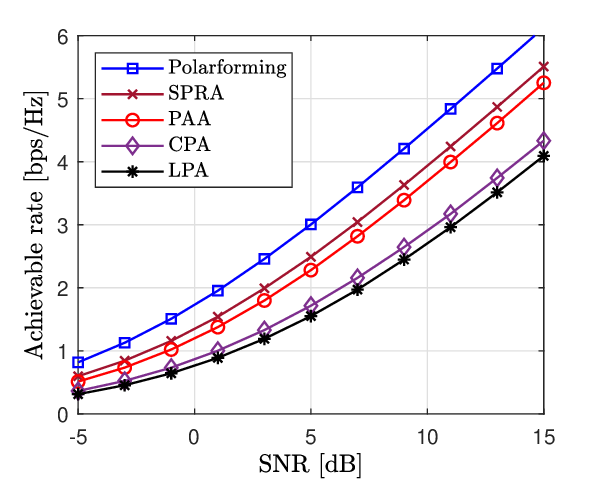}
	\caption{Achievable rate of polarforming versus SNR when it is applied at both the transmitter and receiver.}
	\label{fig_rate_snr}
\end{figure}

Fig.\;\ref{fig_rate_snr} investigates the achievable rates of polarforming applied at both the transmitter and receiver and benchmark schemes versus SNR.   From this figure, it is evident that the proposed scheme achieves superior achievable rates compared to all benchmark schemes, with the performance gap remaining nearly consistent as SNR increases.  Specifically, at the rate of $4$ bps/Hz, the proposed scheme achieves SNR gains of $1.9$ dB, $2.7$ dB, $5.6$ dB, and $6.3$ dB over the conventional SPRA, PAA, CPA, and LPA schemes, respectively, for the considered SISO scenario.  This result demonstrates the significant advantage of polarforming in enhancing communication performance.

\section{Conclusion} \label{sec5}
In this paper, we introduced and investigated the new concept of polarforming that enables the antenna to configure its polarization state to match the polarization of EM waves.  We proposed a novel PS-based PRA and studied a SISO system aided by this new type of PRA.  For the PS-based PRA, we developed an alternating optimization approach to maximize the receive SNR by iteratively tuning the phase shifts at both the transmitter and receiver.  Furthermore, simulation results validated the effectiveness of the PS-based PRA system.  It was shown that PS-based PRAs outperform conventional PRAs and FPAs in combating channel depolarization and adapting to channel fading.

\begin{thebibliography}{00}
	\bibitem{ref_DZJZ25}
	J. Ding, Z. Zhou, B. Jiao, and R. Zhang, ``Polarforming for wireless networks: Opportunities and challenges,'' arXiv preprint \textit{arXiv:2505.20760}, 2025.
	
	\bibitem{ref_ZMZ24}
	L. Zhu, W. Ma, and R. Zhang, ``Movable antennas for wireless communication: Opportunities and challenges,'' \textit{IEEE Commun. Mag.}, vol. 62, no. 6, pp. 114-120, Jun. 2024.
	
	\bibitem{ref_DZL24}
	J. Ding, Z. Zhou, W. Li, C. Wang, L. Lin, and B. Jiao, ``Movable antenna-enabled co-frequency co-time full-duplex wireless communication,'' \textit{IEEE Commun. Lett.}, vol. 28, no. 10, pp. 2412-2416, Oct. 2024.

	\bibitem{ref_LDZ25}
	L. Lin, J. Ding, Z. Zhou, and B. Jiao, ``Power-efficient full-duplex satellite communications aided by movable antennas,'' \textit{IEEE Wireless Commun. Lett.}, vol. 14, no. 3, pp. 656-660, Mar. 2025.

	\bibitem{ref_DZJ25}
	J. Ding, Z. Zhou, and B. Jiao, ``Movable antenna-aided secure full-duplex multi-user communications,'' \textit{IEEE Trans. Wireless Commun.}, vol. 24, no. 3, pp. 2389-2403, Mar. 2025.
	
	\bibitem{ref_WST25}
	K.-K. Wong, A. Shojaeifard, K.-F. Tong and Y. Zhang, ``Fluid antenna systems,'' \textit{IEEE Trans. Wireless Commun.}, vol. 20, no. 3, pp. 1950-1962, Mar. 2021.
	
	\bibitem{ref_DZZ25}
	J. Ding, L. Zhu, Z. Zhou, B. Jiao, and R. Zhang, ``Near-field multiuser communications aided by movable antennas,'' \textit{IEEE Wireless Commun. Lett.}, vol. 14, no. 1, pp. 138-142, Jan. 2025.

	\bibitem{ref_MWN24}
	W. Mei, X. Wei, B. Ning, Z. Chen, and R. Zhang, ``Movable-antenna position optimization: A graph-based approach,'' \textit{IEEE Wireless Commun. Lett.}, vol. 13, no. 7, pp. 1853-1857, Jul. 2024.

	\bibitem{ref_DZS24}
	J. Ding, Z. Zhou, X. Shao, B. Jiao, and R. Zhang, ``Movable antenna-aided near-field integrated sensing and communication,'' arXiv preprint \textit{arXiv:2412.19470}, 2024.

	\bibitem{ref_DSP25}
	Z. Ding, R. Schober, and H. Vincent Poor, ``Flexible-antenna systems: A pinching-antenna perspective,'' \textit{IEEE Trans. Commun.}, early access, Mar. 2025.

	\bibitem{ref_OB23}
	\"O. \"Ozdogan and E. Bj\"ornson, ``Massive MIMO with dual-polarized antennas,'' \textit{IEEE Trans. Wireless Commun.}, vol. 22, no. 2, pp. 1448-1463, Feb. 2023.
	
	\bibitem{ref_XHL14}
	L. D. Xu, W. He, and S. Li, ``Internet of Things in industries: A survey,'' \textit{IEEE Trans. Ind. Informat.}, vol. 10, no. 4, pp. 2233–2243, Nov. 2014.
	
	\bibitem{ref_KRR15}
	J. M. Kovitz, H. Rajagopalan, and Y. Rahmat-Samii, ``Design and implementation of broadband MEMS RHCP/LHCP reconfigurable arrays using rotated E-shaped patch elements,'' \textit{IEEE Trans. Antennas Propag.}, vol. 63, no. 6, pp. 2497–2507, Jun. 2015.
	
	\bibitem{ref_KM15}
	S.-C. Kwon and A. F. Molisch, ``Capacity maximization with polarization-agile antennas in the MIMO communication system,'' in \textit{Proc. IEEE Global Commun. Conf. (GLOBECOM)}, San Diego, CA, USA, Dec. 2015, pp. 1–6.
	
	\bibitem{ref_CH24}
	M. R. Castellanos and R. W. Heath, ``Linear polarization optimization for wideband MIMO systems with reconfigurable arrays,'' \textit{IEEE Trans. Wireless Commun.}, vol. 23, no. 3, pp. 2282-2295, Mar. 2024.
	
	\bibitem{ref_DGM17}
	F. A. Dicandia, S. Genovesi, and A. Monorchio, ``Analysis of the performance enhancement of MIMO systems employing circular polarization,'' \textit{IEEE Trans. Antennas Propag.}, vol. 65, no. 9, pp. 4824–4835, Sep. 2017.
	
	\bibitem{ref_HCS16}
	Y. He, X. Cheng, and G. L. St{\"u}ber, ``On polarization channel modeling,'' \textit{IEEE Wireless Commun.}, vol. 23, no. 1, pp. 80-86, Feb. 2016.
		
\end{thebibliography}
\end{document}